\documentclass[fleqn,usenatbib]{mnras}

\usepackage{newtxtext,newtxmath}

\usepackage[T1]{fontenc}
\usepackage{ae,aecompl}

\usepackage{graphicx}	
\usepackage{amsmath}	
\usepackage{amssymb}	
\usepackage{rotating}
\usepackage{lscape}

\newcommand{\RomanNumeralCaps}[1]
    {\MakeUppercase{\romannumeral #1}}
    
\newcommand{\Paper}[1]
    {Paper \RomanNumeralCaps{#1}}
    
\newcommand{\steel}
    {\textsc{steel} }
    
\newcommand{\LCDM}{$\Lambda$CDM }

\usepackage{etoolbox}
\makeatletter
\patchcmd\@combinedblfloats{\box\@outputbox}{\unvbox\@outputbox}{}{%
  \errmessage{\noexpand\@combinedblfloats could not be patched}%
}%
\makeatother

\title[STEELIII]{The significant effects of stellar mass estimation on galaxy pair fractions.}

\author[]{
Philip J. Grylls,$^{1}$\thanks{E-mail: P.Grylls@soton.ac.uk}
F. Shankar,$^{1}$\thanks{E-mail: F.Shankar@soton.ac.uk}
C. Conselice,$^{2}$\thanks{E-mail: conselice@gmail.com}
\\
$^{1}$Department for Physics and Astronomy, University of Southampton, Highfield SO171BJ, UK\\
$^{2}$University of Nottingham, School of Physics and Astronomy, NG7 2RD, UK
}

\date{Accepted XXX. Received YYY; in original form ZZZ}

\pubyear{2018}

\begin{document}
\label{firstpage}
\pagerange{\pageref{firstpage}--\pageref{lastpage}}
\maketitle

\begin{abstract}
There exist discrepancies in measurements of the number and evolution of galaxy pairs. The pair fraction appears to be sensitive to both the criteria used to select pair fraction and the methods used to analyze survey data. This paper explores the connection between stellar mass estimation and the pair fraction of galaxies making use of $\steel$, the Statistical sEmi-Emprical modeL. Previous results have found the pair fraction is sensitive to choices made when selecting what qualifies as a pair, for example luminosity or stellar mass selections. We find that different estimations of stellar mass such as photometric choice mass-to-light ratio or IMF that effect the stellar mass function also significantly affect the derived galaxy pair fraction. By making use of the galaxy halo connection we investigate these systematic affects on the pair fraction. We constrain the galaxy halo connection using the stellar-mass-halo-mass relationship for two observed stellar mass functions, and the Illustris TNG stellar mass function. Furthermore, we also create a suite of toy models where the stellar-mass-halo-mass relationship is manually changed. For each stellar-mass-halo-mass relation the pair fraction, and its evolution, are generated. 
We find that enhancements to the number density of high mass galaxies cause steepening of the stellar-mass-halo mass relation, resulting in a reduction of the pair fraction. 
We argue this is a considerable cause of bias that must be accounted for when comparing pair fractions.
\end{abstract}




\section{Introduction}
\label{sec:Intro}
$\Lambda$CDM cosmology predicts the hierarchical assembly of dark matter haloes.
Throughout the history of the Universe haloes have grown in mass and size via two pathways. 
Firstly haloes grow via smooth accretion gradually accreting dark matter from the surrounding environment. 
The secondary growth mechanism is via the accretion and gradual absorption of smaller haloes, known as subhaloes. 
After accretion subhaloes survive as substructure of the central/host halo gradually losing mass and sinking to the center of the potential well though dynamical friction.
At early epochs the gas directly follows the collapse of the dark matter sinking to the bottom of the potential wells cooling and forming stars \citep{Mo1998TheDiscs}. Larger dark matter haloes had a stronger gravitational influence creating a deeper potential well capturing more gas, and are thus expected to be associated, on average, with larger galaxies. The efficiency with which baryons form galaxies in haloes has been shown to depend on the halo mass, with smaller galaxies growing more with increasing halo mass than larger galaxies. The latter effect is seen in the shape of the stellar-mass-halo-mass (hereafter, SMHM) relation, where at a particular stellar/halo mass the slope of the relationship changes \citep[e.g.][]{Shankar2006NewFormation, Moster2010, Behroozi2013THE0-8, Moster2018Emerge10, Shankar2014,  Conselice2018TheApproach, Grylls2019AClusters, Grylls2019PredictingSteel.}.

Frequent or massive mergers are thought to induce morphological changes in galaxies. Galaxies, after experiencing a massive merger, where the minor galaxy is at least a quarter of the mass of the central galaxy, are thought to lose their disk-like morphology and transform into elliptical galaxies \citep{Negroponte1983SimulationsGalaxies, DeLucia2006TheGalaxies}. For this reason it is important to understand the frequency and nature of mergers between galaxies to achieve a complete and coherent picture of galaxy formation and evolution. However, galaxy mergers occur on gigayear timescales and therefore it is not possible to directly observe the rate or consequence of galaxy mergers. However, by inferring the merger rate over longer timescales it can be used as a powerful constraint on cosmological parameters \citep{Conselice2014GalaxyParameters}
The traditional approach to estimate a measure of galaxy mergers is to instead count galaxy pairs at a given separation, and then assign a merging timescale to infer the rate of galaxy merging \citep{Conselice2003A3,Conselice2008TheField,Mundy2017A3.5,Duncan2019ObservationalFields}. However, the approach of counting pairs is complicated by systematic differences when selecting galaxies, for example the evolution of the pair fraction appears to change if a selection is made by flux ratio or made by stellar mass ratio \citep{Man2016RESOLVING03}.

In \citet{Grylls2019AClusters} (Hereafter \Paper{1}) we introduced a the Statistical sEmi-Empirical modeL $\steel$. \steel and other transparent semi-empirical approaches, have had multiple successes:
\begin{itemize}
    \item Mapping galaxies into dark matter haloes using the SMHM relation, derived from the relative abundance of dark matter haloes and galaxies.
    \item Following the mergers of the underlying host dark matter halos.
    \item Computing the implied rate of galaxy mergers.
\end{itemize}
Uniquely in \steel these properties do not follow discrete dark matter merger histories, `merger trees', instead they are calculated using a statistical dark matter accretion history giving access to unbiased population averages. In \Paper{1} we use this to calculate the distributions of satellite galaxies in massive haloes and in \citet{Grylls2019PredictingSteel.} (Hereafter \Paper{2}) this is extended to high redshift. In \Paper{2} a comparison is made with observations of satellite distributions and stellar mass functions over multiple epochs, such we are able to confirm that the satellite distributions generated by \steel are consistent with observations over a large range of redshifts. As \steel is consistent with satellite distributions over multiple epochs by extension the satellite accretion histories are a reliable estimate of the true accretion.

The ability to reliably predict the galaxy merger rate is particularly powerful as it provides an additional tool to test the \LCDM cosmological model. The galaxy merger rate is intrinsically connected to the dark matter assembly so that galaxies can be used as a proxy for the dark matter structure. However, there are notable systemics that could affect the reliability of using the galaxy merger rate from models. The primary tool in this analysis is the SMHM relation which is heavily dependent on the shape of the input stellar mass function, the comoving number density of galaxies of a given stellar mass. The stellar mass function is affected by several observational systematics, notably, choice of mass-to-light ratios, stellar initial mass functions, light profiles e.t.c... In the last decade it has been shown that the stellar mass function is significantly higher in the high mass end than previously thought, this is due to a number of improvements in stellar mass calculation including, complete fitting of the extended light profile\citep{Bernardi2016TheEvolution,Bernardi2017TheProfile}), S\`ersic + exponential fitting models \citep{Meert2015ASystematics} and improved sky subtractions. Using better stellar mass estimates the number density of high mass galaxies is increased, this in turn steepens the high mass slope of the SMHM relationship \citep{Shankar2014ON1,Kravtsov2018StellarHalos}. 

The effect of the stellar mass functions and the systematics introduced into the SMHM relation were investigated in \Paper{2}. It was shown that shallow SMHM relations, associated with the previous stellar mass estimations, coupled to hierarchical \LCDM cosmology create satellite accretion histories that are incompatible with central galaxy growth. In contrast stellar mass functions with enhanced high mass number density predict central galaxies grow more rapidly with cosmic time, such higher growth rates were found to be consistent with the total satellite accretion predicted by \LCDM cosmology. Given the pair fraction is used as an observational estimate of satellite accretion through galaxy mergers, one would expect that where stellar mass estimations impact the accretion rates they should also produce a difference in modeled pair fractions. 

In this work we investigate how the observed pair fraction changes systematically with varying SMHM relationship. We use $\steel$ which is capable of producing state-of-the-art satellite abundance mocks using a statistical dark matter accretion history and the galaxy-halo connection. We show that a well designed and flexible Semi-Empirical model should be used as an essential analytic tool for understanding how observational modelling assumptions, such as the estimation of stellar mass, may propagate in unpredictable ways. 

This paper is laid out as follows. In Section \ref{sec:Data} we describe the comparative simulation data used. In Section \ref{sec:Method} we summarize \steel the STastical sEmi-Empirical modeL and the extensions added for the analysis in this work. In Section \ref{sec:Results} we show a systematic analysis of how the SMHM relation effects the pair fraction then compare the output of \steel  using altered SMHM relations to match simulation and observational results to show the magnitude of the differences. In Sections \ref{sec:Discussion} \& \ref{sec:Conclusions} we situate our results in a wider context and conclude. 

\section{Data}
\label{sec:Data}

\subsection{Stellar Mass Functions}
\label{subsec:SMF}
The input SMHM relations used in \steel are constrained using observed stellar mass functions. \steel uses two stellar mass functions at redshift $z = 0.1$, each using the Sloan Digital Sky Survey Data Release 7 \citep{Abazajian2009THESURVEY, Meert2015ASystematics}, these are made using a S\`ersic-Exponential fit `PyMorph' \citep{Meert2015ASystematics, Bernardi2016TheEvolution} and a de Vaucoulers fit `cmodel' \citep{deVaucouleurs1948RecherchesExtragalactiques}. These stellar mass functions are representative of the previous photomotries (cmodel) and the enhanced high mass end (PyMorph) described in Section \ref{sec:Intro}. To constrain the SMHM relations at higher redshift (0.3 < z < 3.3) we use the stellar mass functions from COSMOS2015 catalogue \citep{Davidzon2017TheSnapshots}. These stellar mass functions are created using \citet{Bruzual2003Stellar2003} stellar population synthesis models. These stellar mass functions can be used directly with the cmodel fits, however a +0.15 dex correction is required to compare to the PyMorph light profile fitting \citep{Mendel2014ASURVEY,Bernardi2013TheProfile}. This correction brings the redshift $z=0.37$ stellar mass function in agreement with the PyMorph $z=0.1$ stellar mass function matching the finding that the stellar mass function does not evolve significantly up to redshift $z = 0.5$ \citep{Bernardi2016TheEvolution}. The differences in the SMHM relations from these fits are discussed in Section \ref{subsec:InfluenceofSMHM} and fitting the parameters given in Section \ref{subsec:SMHM}.

\subsection{Illustris}
We use data extracted from the Illustrius TNG simulation JupyterLab public data mirror \citep{Springel2018FirstClustering, Nelson2018TheRelease}. Illustrius TNG is a state of the art, large volume, cosmological, gravo-magnetohydrodynamical simulation. To make comparisons to the results from \steel we utilize the group and subhalo galaxy/dark matter catalogues to explore the distributions of galaxy pairs in group and cluster environments from Illustrius TNG.

\section{Method}
\label{sec:Method}

In this paper we highlight difficulties present when comparing models to observations, specifically pair fractions. 
Foremost, the difficulties stem from the assumptions inherent in a given stellar mass estimation. 
For example, using the Sloan Digital Sky Survey Data Release 7 (SDSS-DR7), two notably different stellar mass functions have been derived. 
Using the cmodel fit \citep{Abazajian2009THESURVEY} produces a stellar mass function with a sharp high mass cutoff, whereas, using a S\'ersic + exponential model \citep{Meert2015ASystematics, Bernardi2016TheEvolution, Bernardi2017ComparingLight} produces a less sharp cutoff and more high mass galaxies. 
The propagation of these differences into the SMHM relation, in conjunction with hierarchical assembly predicted by \LCDM cosmology, creates different galaxy assembly histories within the same cosmology.
An empirical model, such as $\steel$, is ideally suited to understand these effects.

By design, an empirical model recreates the stellar mass function at all redshifts, something other modeling techniques have historically struggled with \citep{Asquith2018CosmicModels}. 
The reproduction of the stellar mass function at all redshifts is essential to derive reliable assembly histories. 
For example, a stellar mass function at high redshift with an excess of low mass galaxies must undergo a prolonged phase of over-merging to match the stellar mass function at low redshift. 
Additionally, \steel has an advantage over more traditional models. Using a statistical dark matter accretion history, we replace discrete dark matter volumes to simulate without volume or resolution constraints. 
In \Paper{1} and \Paper{2} we showed that using the S\'ersic + exponential fits, \steel recreates the distributions of satellite galaxies over a large range of cosmic epochs.
In \Paper{2} we showed the in-situ vs ex-situ growth ratios for central galaxies for both stellar mass functions. We found that flatter SMHM relations create a satellite galaxy accretion history that is too fast with respect to the predicted growth of the central galaxy. 
As observations of galaxy pair fractions are often used to create an estimation of the galaxy merger rate \citep{Mundy2017A3.5, Mantha2018Major0}. It follows that the discrepancies found in the satellite mass accretion introduced by the SMHM relation should also alter the galaxy pair fraction.

In this Paper we discuss the updates to \steel that enable the extraction of a galaxy pair fraction given a choice of underlying stellar mass function. Using a toy model we perturb in isolation each parameter of the SMHM relation described in Section \ref{subsec:SMHM}. We show that shape of the SMHM relationship directly affects the normalisation and evolution of the expected galaxy pair fraction. 

In this section we begin by describing how the SMHM relation propagates into the pair fraction with the aid of two cartoons in Section \ref{subsec:InfluenceofSMHM}. We provide the quantitative fits for the SMHM relations used in this work in Section \ref{subsec:SMHM}. We then describe the statistical accretion history essential to \steel in Section \ref{subsec:StatAccHist}. Finally, we describe the method used to apply a spacial distribution to the statistical satellite population in Section \ref{subsec:Seperation}.

\subsection{Influence of the SMHM relation}
\label{subsec:InfluenceofSMHM}
The photometric choice, or other systematics such as mass-to-light ratio, the initial mass function, or background subtraction, used to calculate galaxy stellar mass, all have a direct impact on the resulting shape of the stellar mass function. For a fixed cosmological model, i.e. the number and distribution of haloes are invariant, changes in the stellar mass function directly correlate to changes in the SMHM relation. The SMHM relation can therefore be used as an effective proxy for the shape of the underlying stellar mass function.

\begin{figure*}
	\centering
	\includegraphics[width = \linewidth]{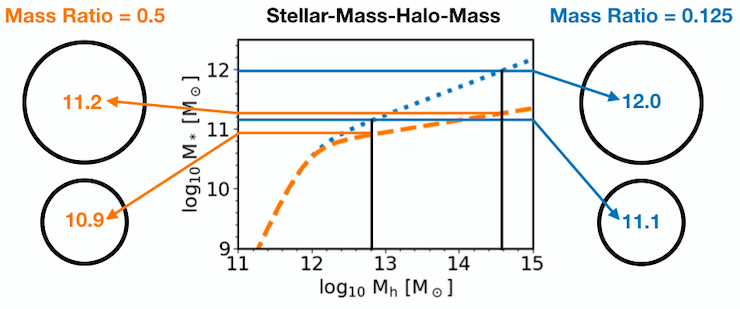}
    \caption{A diagram showing the how the SMHM relation can impact the stellar mass ratio of galaxies mapped into identical halos. The black vertical lines represent the (fixed) halo masses that are seeded with stellar mass. Coloured horizontal lines are drawn at the intersection of the halo masses and the SMHM relation these show the stellar mass seeded by each relation also shown by the numbers in matching colour on either side of the plot.
    The steeper SMHM relation (blue) creates a smaller stellar mass ratio as the change in halo mass maps to a much larger stellar mass difference.}
	\label{fig:MassRatioCartoon}
\end{figure*}

In this work we show that different SMHM relations generate distinct pair fractions and merger rates. 
Stellar mass functions with greater number densities of high-mass galaxies, naturally map larger galaxies into smaller haloes due to their higher relative abundances, resulting in steeper high-mass slopes for the SMHM relations.
In Figure \ref{fig:MassRatioCartoon} we show an illustrative cartoon of how different SMHM relations affect the galaxy mass ratios. 
For two identical mass halo pairs we see that a SMHM relation with a steeper slope causes a substantial difference in the stellar mass ratio mapped into the halo pairs. 
In general, a shallower relation causes many more massive pairs than a steeper relation.

\begin{figure*}
	\centering
	\includegraphics[width = \linewidth]{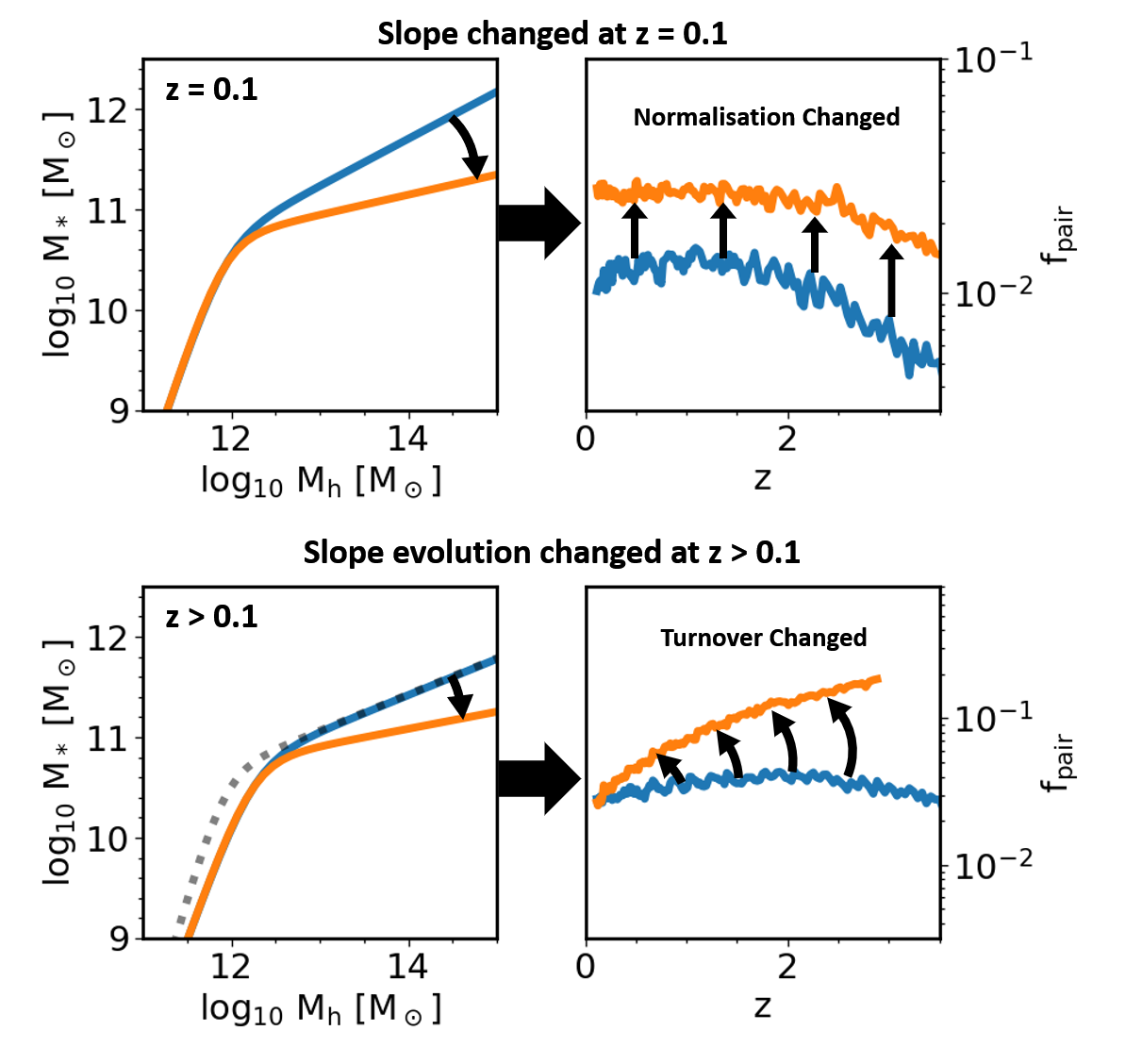}
    \caption{A diagram showing the how the SMHM relation can impact the pair fraction. The top row shows how reducing the high mass slope of the SMHM relation increases the number of pairs at all redshifts. The bottom row shows the redshift $z=0$ relation as a grey dotted line, two relations at redshift where the relation is not evolved or evolved to be shallower are shown in blue and orange respectively. For this evolving SMHM relation the pair fractions are found to increase. In each case the reason for the increase can be explained by referencing Figure \ref{fig:MassRatioCartoon} where making the relation shallower seeds more massive pairs.}
	\label{fig:SMHM_PF_Cartoon}
\end{figure*}

In Figure \ref{fig:SMHM_PF_Cartoon} a cartoon is shown to give an example of the expected difference in the pair fraction when changing the SMHM relation.
We define pair fraction as the fraction of galaxies of a given mass that have a companion with a mass equal to or greater than a quarter of the primaries mass within 5-30 kpc. 
The left hand column shows the SMHM relations and the right column the implied pair fractions and their evolution with redshift. 
In the top row we compare a steep high-mass slope to a flatter slope, where the slope has been changed at redshift $z = 0.1$.
In the bottom row we compare an evolving and non-evolving high-mass slope.

The steepening of the high-mass slope increases the number of pairs created and the normalisation of the pair fraction increases. 
In the bottom row we show the effects of having a slope that flattens at higher redshift. 
We show the redshift $z=0.1$ relation in grey and the relations with the unchanged and changed slopes in blue and orange respectively. 
The main effect of varying the evolution of the high-mass slope in the SMHM relation is to change the behaviour of pair fraction with redshift. 
A steeper slope tends to turn over the pair fraction and vice-versa.
The behaviours reported in Figure \ref{fig:SMHM_PF_Cartoon} are what one would expect given Figure \ref{fig:MassRatioCartoon}, where shallower slopes give higher fractions. 
Furthermore, from Figure \ref{fig:SMHM_PF_Cartoon} (and from Figure \ref{fig:PairFracSystematic}, shown in Section \ref{sec:Results}), it can be concluded that almost any pair fraction time dependence could be produced by appropriately altering the input SMHM relation. 
It is relevant to stress here that relatively minor changes in the stellar mass function can cause qualitative differences in the SMHM relation and, by extension, in the shape and normalization of pair fractions at any cosmic epoch.

\subsection{SMHM relation}
\label{subsec:SMHM}
A double power-law relation similar to \citet{Moster2010} is used to parameterise the SMHM relation. The parameters M, N, $\beta$, and $\gamma$ control, respectively, the position of the knee, the normalization, the low-mass, and the high-mass slope at redshift $z=0.1$. Each parameter has an associated redshift evolution factor given by:

\begin{equation}
\label{eqn:MosAbn}
\begin{split}
M_*(M_h, z) &= 2M_hN(z)\Big[ \Big( \frac{M_h}{M_{n}(z)}\Big) ^{- \beta(z)} + \Big( \frac{M_h}{M_{n}(z)}\Big)^{\gamma(z)} \Big ]^{-1}\\
N(z) &= N_{0.1} +N_z\Big(\frac{z-0.1}{z+1}\Big)\\
M_{n}(z) &= M_{n,0.1} +M_{n,z}\Big(\frac{z-0.1}{z+1}\Big)\\
\beta(z) &= \beta_{0.1} +\beta_z\Big(\frac{z-0.1}{z+1}\Big)\\
\gamma(z) &= \gamma_{0.1} +\gamma_z\Big(\frac{z-0.1}{z+1}\Big).
\end{split}
\end{equation}
We use the abundace matching fits from \Paper{2}, shown in Table \ref{tab:AbnResult}. The parameters for Equation \ref{eqn:MosAbn} are given for a S\'ersic-Exponential fit stellar mass function (PyMorph) and a de Vaucoulers fit stellar mass function (cmodel). The differences between these two fits are given in Section \ref{subsec:SMF}. In brief, PyMorph gives more high mass galaxies and results in a steeper high mass slope of the SMHM relation. Conversely cmodel has a lower number of high mass galaxies and creates a flatter high mass slope. For completeness, we also include a SMHM relation that well fits the outputs of the Illustris TNG which is steeper than PyMorph.                                      

\begin{table*}
\centering
\begin{tabular}{l|llllllll}
        & M\_n & N     & $\beta$ & $\gamma$ & $M_{n,z}$ & N\_z   & $\beta_z$ & $\gamma_z$ \\ \hline
\\
cmodel  & $11.91_{-0.34}^{+0.40}$ & $0.029_{-0.013}^{+0.018}$ & $2.09_{-1.02}^{+1.21}$    & $0.64_{-0.10}^{+0.11}$     & $0.52_{-0.19}^{+0.24}$       & $-0.018_{-0.004}^{+0.005}$ & $-1.03_{-0.34}^{+0.049}$     & $0.084_{-0.14}^{+0.20}$     \\
\\
PyMorph & $11.92_{-0.36}^{+0.39}$ & $0.032_{-0.012}^{+0.016}$ & $1.64_{-0.73}^{+0.85}$     & $0.53_{-0.11}^{+0.11}$     & $0.58_{0.19}^{+0.15}$        & $-0.014_{-0.006}^{+0.007}$ & $-0.69_{-0.36}^{+0.29}$      & $0.03_{-0.147}^{+0.154}$    \\
\\
TNG & $11.8$ & $0.018$ & $1.5$ & $0.31$ & $0.0$ & $-0.01$ & $0$ & $-0.12$   
\end{tabular}
\caption{The SMHM relation fits for cmodel, PyMorph, and TNG. For the cmodel and PyMorph data, the errors are the 16th and 86th percentile from the MCMC fiting. }
\label{tab:AbnResult}
\end{table*}                                               

\subsection{Statistical Accretion Histories}
\label{subsec:StatAccHist}
Traditional simulations such as Hydrodynamical, Semi-Analytic, or Semi-Empirical models, simulate the dark matter background of the universe using a cosmological box or a discrete set of merger trees. 
Both cosmological boxes and merger trees simulate a discrete cosmological volume, and in any given volume there will be a limited number of massive haloes. 
Due to the significant decrease in number density with increasing halo mass, haloes of even a couple of orders of magnitude smaller than the most massive halo in the simulation are found significantly more frequently.
Due to the large difference in number density mergers between haloes of similar mass are extremely rare, especially compared to halo mergers with haloes with a low mass ratio.
\steel removes the dependence on discrete halo sets by using a `statistical accretion history': haloes and mergers of any mass and mass ratio are simulated equally regardless of number density. 

The full method for creating the `statistical accretion history' is given in \Paper{1} and a summary of the method is provided in \Paper{2}. In brief, we start from the average growth history of central haloes and at each epoch compute the unevolved subhalo mass function\footnote{This is the total number density of subhaloes accreted by a central halo over its growth history as a function of the mass ratio $M_{sat}/M_{cen}$ where the satellite masses are frozen at infall.}. At each time step the growth of the unevolved subhalo mass function is attributed to accretion of new subhaloes to create the accreted subhalo mass function (simply the number density of a given subhalo mass accreted at a given epoch for a given central mass history). Each bin of the accreted subhalo mass function is assigned a dynamical time, in essence the time to merge with the central galaxies. The surviving subhalo mass function\footnote{The total number density of subhaloes that one would expect to still be present in the parent halo at a given epoch.} is created by summing each mass of accreted subhaloes at each time step that have not yet merged with their centrals by the redshift of observation. In this work we use the PyMorph abundance matching fits from \Paper{2} to associate each bin of the  accreted subhalo mass function with a distribution of galaxies at infall. The satellite galaxies associated to subhalo bins that have reached the end of their dynamical times are summed to contribute to the average galaxy accretion history for a given dark matter halo.

\subsection{Galaxy Separations}
\label{subsec:Seperation}
Calculation of the pair fraction requires an estimate of the distance between the central galaxy and the satellite galaxy, as we rely on our statistical accretion history, and do not have discrete halos, we assign each subhalo bin an average distance to the central galaxy. The subhaloes start at the viral radius of the central halo. The distance to the centre then reduces proportionally to the amount of dynamical time remaining \citep{Guo2011FromCosmology}. Throughout this work unless otherwise stated the pair fraction is calculated as the number density of satellites with a mass ratio of above 1/4 within 5 to 30 kpc of the central galaxy divided by the total number of central galaxies within the mass selection.

\section{Results}
\label{sec:Results}
In this section we test the impact of two different photometric choices, PyMorph \citep{Meert2015ASystematics} and cmodel \citep{Abazajian2009THESURVEY}, on the pair fractions. Fixing the dark matter halo assembly and cosmology, the two photometries generate two distinct SMHM relations where the main difference lies in the high mass slope. 
To gain deeper insight into how the input SMHM relation propagates into the pair fractions, in Section \ref{subsec:SysAna} we show a toy model where each parameter of the SMHM relation is perturbed independently. Analysis of this toy model will inform which SMHM relation parameters affect aspects of the pair fraction, applying this in Section \ref{subsec:SimObsRes} we then attempt to recreate pair fractions found in the Illustris TNG simulation and in data from \citet{Mundy2017A3.5} by manipulating the SMHM relationship. 
Finally, using the SMHM relation that best fits the observed pair fraction we compute satellite accretion histories following the methods in \Paper{2}. 
In brief, the amount of stellar mass growth expected from satellite accretion is compared to the central stellar mass growth predicted by abundance matching. If the satellite accretion is greater than the central growth the SMHM relation is non-physical and the stellar mass estimation used for the pair fraction must be questioned.

In this section we investigate the distribution of satellites around central galaxies when using different SMHM relations. This distribution is primarily dominated by the halo substructure, for this reason it is essential to make sure our selection criteria for galaxies always returns the same halo population. As we actively change the stellar masses mapped into any given halo mass one cannot use a stellar mass cut to achieve this result. From a simulation where the haloes are known, one could simply select by halo mass, however, to better match observation where haloes are not known we make a constant number density selection. A constant number density selection will always return objects that share a given number density and are therefore associated to the same haloes as we are taking the halo structure as a fixed quantity. A similar technique was employed by, \citet[e.g.][]{Leja2013TRACINGSELECTION, Mundy2015Tracing3} to trace the evolution of galaxy populations; galaxies at high redshift with a given comoving number density are assumed to be the progenitors of later populations with the same abundance. 
In this work we use a central stellar mass selection from the PyMorph stellar mass estimation of $M_{*} = 10^{11} - 10^{11.6} M_{\odot}$ (or $10^{9.5} - 10^{10.1} M_{\odot}$ when considering the low-mass slope controlled by $\beta$). The number-density range of the aforementioned mass cut is then computed and galaxies are selected from the other SMHM relations which share this number density.
An example of this selection can be seen in Figure \ref{fig:PairFracSystematic}, the shaded horizontal band shows the stellar masses for each SMHM relation that share number density.

\subsection{Systematic Analysis}
\label{subsec:SysAna}

We use the flexible nature of \textsc{steel} to create a toy model where each of the main parameters (M, N, $\beta$, $\gamma$), and their evolutionary factors (M$_z$, N$_z$, $\beta_z$, $\gamma_z$), governing the SMHM relation are adjusted in turn to explore the affect on the galaxy pair fractions. Table \ref{tab:PairFracSysInput} details the change made to the SMHM relation for each parameter. 

Figure \ref{fig:PairFracSystematic} shows each of the SMHM relations in the outer four panels, the reference SMHM relation PyMorph is shown in blue at redshifts $z = 0.1$ (dotted line) and $z = 2$ (dashed line) in each panel, the modified redshift $z = 0.1$ relation is then shown in orange, and the increased and decreased (dashed red and green) evolution are shown at redshift $z = 2$. The inner four panels follow the same colour convention. When changing M, the knee parameter, a large increase in the pair fraction is found from a lower knee: The shallower high mass slope is extended therefore more haloes are seeded in the mass range for pairs. We see the same effect at high redshift, the lower value of M at high redshift creates a higher pair fraction. 

The normalization parameter, N, creates little change in the pair fraction as expected because the mass ratios are largely unaffected. The low mass slope parameter, $\beta$, affects the seeding of smaller galaxies hence a lower mass range is used for the consistent number density cut. Due to the steepness of the low mass slope the fraction of pairs is lower in this mass cut.
 Finally, when the high mass slope parameter, $\gamma$, is altered more pairs are found at high and low redshift when the slope is shallow. This is again attributed to more galaxies seeded within the mass ratio range.

\begin{table}
\centering
\caption{The adjustments to the SMHM relation used in Figure \ref{fig:PairFracSystematic}.}
\label{tab:PairFracSysInput}
\begin{tabular}{|c|cccc|} \hline
             & PyMorph   & $X_{0.1, alt}$  & $X_{z, +}$  & $X_{z, -}$  \\ \hline
$M$          & 11.92 & -0.25 & -     & -     \\ 
$M_{z}$      & 0.58   & -     & +0.1  & -0.1  \\ \hline
$N$          & 0.032 & +0.04 & -     & -     \\
$N_{z}$      & -0.014 & -     & +0.007 & -0.007 \\ \hline
$\beta$      & 1.64  & -0.3  & -     & -     \\
$\beta_{z}$  & -0.69  & -     & +0.3  & -0.3  \\ \hline
$\gamma$     & 0.53  & +0.06 & -     & -     \\
$\gamma_{z}$ & -0.03  & -     & +0.2  & -0.2  \\ \hline
\end{tabular}
\end{table}

\begin{landscape}
\begingroup
\begin{figure}
	\centering
	\includegraphics[width = \linewidth]{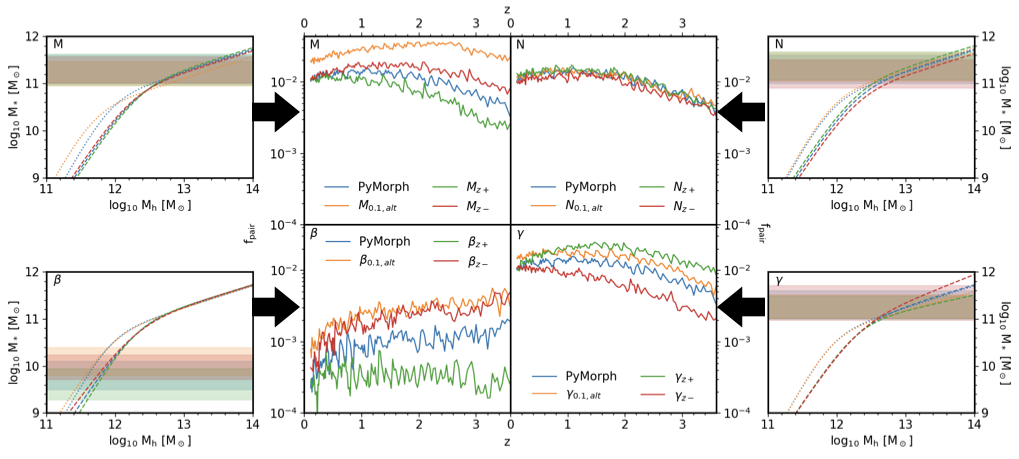}
    \caption{Each of the panel pairs (M, N, $\beta$, $\gamma$) shows the input SMHM relation in the outer plot and the modelled pair fraction evolution in the center plot. Each pair investigates adjustments to the given parameter of the SMHM relation (M, N, $\beta$, $\gamma$). Each pair shows the reference SMHM relation `PyMorph' in blue, the relation adjusted at redshift $z = 0.1$ keeping the same SMHM relation evolution parameters in yellow. The red and green lines respectively have the evolution parameter altered such that the evolution parameter is increased or decreased with respect to the PyMorph relation from S\`ersic Exponential fits \citep{Meert2015ASystematics}. In the outer (SMHM relation) plots dotted lines are $z = 0.1$ relations and dashed lines are $z = 2$ relations the PyMorph is shown at both epochs for comparison. Finally the shaded bands in the outer plots show the consistent number density selections used in the center plots.}
	\label{fig:PairFracSystematic}
\end{figure}
\endgroup
\end{landscape}

\subsection{Simulation and Observational Results}
\label{subsec:SimObsRes}
The observed pair fraction is known to have discrepancies based on the galaxy property used to calculate the ratio. In \citet{Man2016RESOLVING03} it is shown that selecting pairs by flux ratio or stellar mass creates differences in the pair fraction evolution. 
In Figures \ref{fig:MassRatioCartoon} \& \ref{fig:SMHM_PF_Cartoon} we show the predicted effect of the determination of stellar mass on the SMHM relation and the propagation of these changes into the pair fraction. 
Through the use of a toy model in Figure \ref{fig:PairFracSystematic} we show how isolated perturbations to the eight SMHM relation parameters propagate into the galaxy pair fraction.
Given this analysis we find that any observation of the pair fraction must be understood in terms of its implicit observational assumptions. 
Furthermore, direct comparison of pair fraction results should only be undertaken under identical stellar mass derivation assumptions, where this is not the case the influence of any differences must be accounted for.
In this section we fit, by making use of $\steel$, observed pair fractions using small changes to the SMHM relation.
We anticipate this modelling can be used to provide corrections to pair fraction results to allow for fair comparisons.

\begin{figure*}
	\centering
	\includegraphics[width = \linewidth]{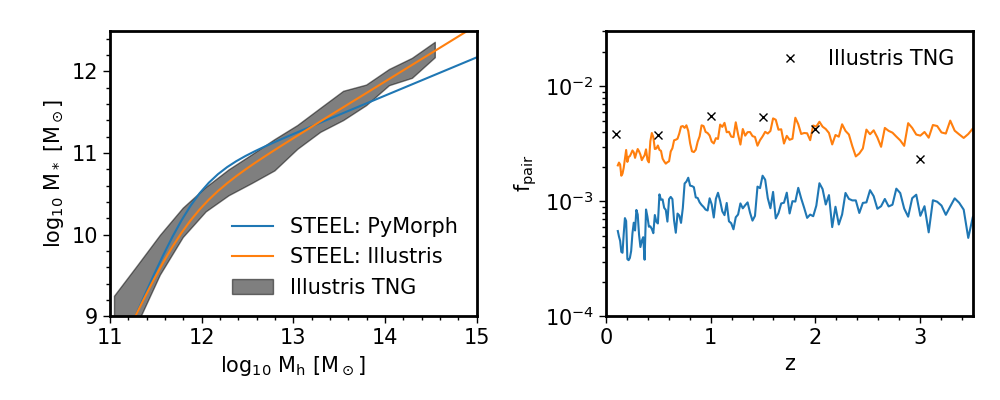}
    \caption{Left: Two SMHM relations are shown from \steel using parameters designed to reproduce the SMHM relation found in the Illustrius TNG simulation (Orange line) and the PyMorph(S\`ersic Exponential) fit parameters (Blue line). The shaded region is the output from the Illustris TNG simulation. Right: The pair fraction, for galaxies in the mass range $M_*$ = $10^{10}M_{\odot}$ to $10^{10.6}M_{\odot}$ generated from \steel is shown for runs using both the SMHM relations, lines follow the same colours as the left hand panel. The pair fraction extracted directly from the Illustris TNG simulation is shown using black crosses.}
	\label{fig:PairFractionIll}
\end{figure*}

In Figure \ref{fig:PairFractionIll} we show the simulated galaxy pair fractions for galaxies in the mass range $M_*$ = $10^{10}$ to $10^{10.6}$. The pair fraction is shown for two different SMHM relation inputs to $\steel$. In blue we show the PyMorph (S\`ersic Exponential) input used as the baseline in Figure \ref{fig:PairFracSystematic}, in orange the input calibrated to match the Illustris TNG simulation. In the right hand panel we see that the the prediction from \steel with the TNG-calibrated input is in good agreement to the pair fraction extracted directly from the Illustris TNG simulation. The pair fraction predicted using the PyMorph input is 0.5 dex lower, this is to be expected as in the mass range we are considering the Illustris TNG simulation SMHM relation is shallower and more pairs are therefore created in a greater mass range of halo mergers. 

\begin{figure*}
	\centering
	\includegraphics[width = \linewidth]{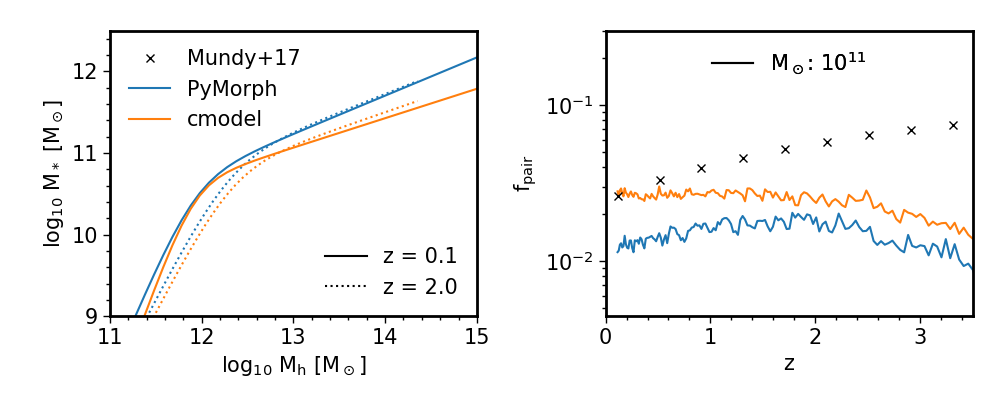}
    \caption{Left: The stellar mass halo mass relations derived from PyMorph (blue) and cmodel (orange) at redshifts 0.1 (solid lines) and 2.0 (dotted lines). Right: The pair fraction evolution for galaxies using both SMHM relations. We make mass cuts, $>10^{10}M_{\odot}$ (dashed line) and $>10^{10}M_{\odot}$ (solid line), in PyMorph and cmodel. The black crosses show the corresponding best fits for the $>10^{11}M_{\odot}$ mass cut from \citet{Mundy2017A3.5}.}
	\label{fig:PairFractionData}
\end{figure*}

Figure \ref{fig:PairFractionData} shows the predicted pair fraction evolution using the two SMHM relations from PyMorph and cmodel presented in Section \ref{subsec:SMHM}. The left panel shows each SMHM relation at redshift $z = 0.1$ and $z = 2.5$. Following the systematic investigation in Figure \ref{fig:PairFracSystematic} we attribute the 0.1 dex difference in pair fraction to the difference in high mass slope between PyMorph and cmodel. The best-fit relation from \citet{Mundy2017A3.5}, shown as black crosses, rises rather than falling as seen from PyMorph and cmodel. We see in Figure \ref{fig:PairFracSystematic} a SMHM relation with a high mass slope that decreases with redshift creating a pair fraction evolution of this nature.

\begin{figure*}
	\centering
	\includegraphics[width = \linewidth]{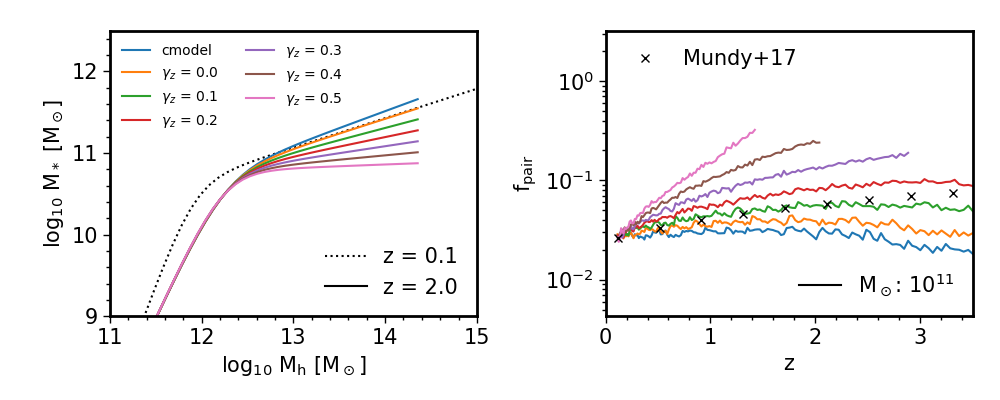}
    \caption{Left: The stellar mass halo mass relations derived from cmodel (black) at redshift $z = 0.1$ (dotted lines) and at $z = 2.0$ (coloured lines) with altered high mass slope evolution parameter. Right: The pair fraction evolution for galaxies each altered SMHM relation. The black crosses show the corresponding best fits for the $>10^{11}M_{\odot}$ mass cut from \citet{Mundy2017A3.5}.}
	\label{fig:PairFractionHMevo}
\end{figure*}

To attempt to match the \citet{Mundy2017A3.5} pair fractions we begin using the cmodel SMHM relation which gives the closest match in pair fraction at low redshift. Following the analysis of Figure \ref{fig:PairFracSystematic} where higher $\gamma_{z}$ increases the pair fraction at high redshift we alter the parameter from 0.0 to 0.5 in steps of 0.1. In Figure \ref{fig:PairFractionHMevo} the left panel shows the SMHM relation at redshift 0.1 as a black dotted line then coloured lines show the relation at redshift $z = 2$ given the different $\gamma_{z}$ parameters. The right panel shows the impact of this evolution on the pair fraction, as predicted higher $\gamma_{z}$ increases the pair fraction with redshift and a value of above 0.1 removes the turnover. Comparing to \citet{Mundy2017A3.5} we see a value of $\gamma_{z}$ between 0.1 and 0.2 best reproduces the rise in pair fraction. 

\begin{figure*}
	\centering
	\includegraphics[width = \linewidth]{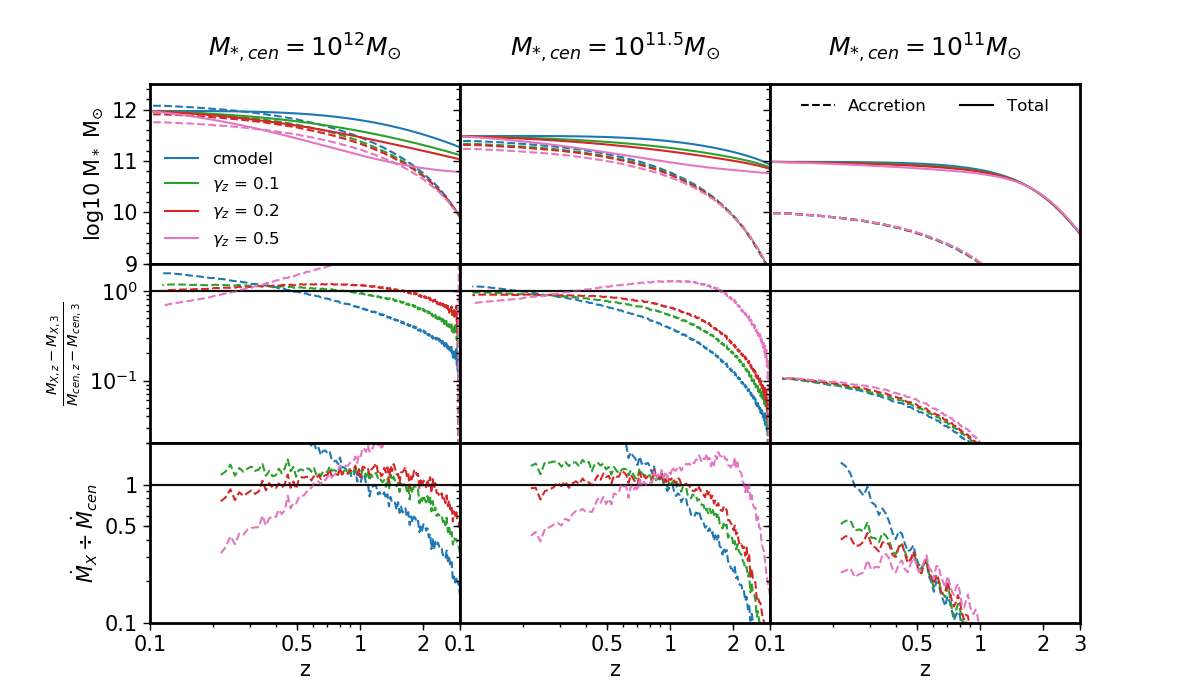}
    \caption{Average `mass tracks' are shown which have central galaxy masses at redshift $z = 0.1$ of $M_{*,cen}$ = $10^{12}$, $10^{11.5}$, and $10^{11}$ $[M_{\odot}]$ from left to right. The satellite galaxy accretion is shown for evolved satellites with a dashed line. The top panels show the total mass of the central (solid lines) and the total mass gained from accretion. The middle panels show the fraction of the total galaxy mass formed from satellite accretion since redshift $z=3$. The bottom panels show the ratio of the mass accretion rate from satellite galaxies to the mass growth rate of the central galaxy predicted by abundance matching. The black horizontal lines in the second and third rows are at unity. The colours are coded to the high mass slope evolution parameter as shown in the legend.}
	\label{fig:SatAcc}
\end{figure*}

In \Paper{2} we show not all SMHM relationships are internally self-consistent within a given \LCDM assembly. Following the average growth of a halo mass bin one can use the SMHM relationship to predict the average galaxy growth for this population of haloes. The average satellite accretion should then be less than, or at most equal to, the central stellar mass growth both in total accretion and instantaneous rate. In Figure \ref{fig:SatAcc} we show the self-consistency in terms of stellar mass for four fits.  We select galaxy populations at three stellar masses as indicated by the column labels.
The galaxies are chosen at redshift $z = 0.1$ and by are followed along the average halo growth histories. 
Each halo growth history is mapped to an average stellar mass history using abundance matching.
For each evolution history \steel calculates the satellite accretion over the galaxy/haloes' history. 

The four fits shown are, the cmodel fit, then the two $\gamma_z$ models that are closest to the \citet{Mundy2017A3.5} pair fraction, and the maximum slope evolution tested $\gamma_z$ = 0.5.
The top row shows the total mass predicted by abundance matching (solid lines) and from satellite accretion (dashed lines), the middle row shows the ratio of mass accreted to total mass gained since redshift $z = 3$, the bottom row shows the ratio of instantaneous satellite accretion to instantaneous growth rate. In the middle and bottom rows the solid black lines are at unity and a model that goes above this line is nonphysical as more matter would have been accreted than the galaxy growth history can account for. From \Paper{2} we know cmodel to be internally inconsistent, for $\gamma_{z}$ = 0.5 the slow growth at redshift $z = 2$ means satellite accretion is far more rapid than galaxy growth, the  $\gamma_{z}$ = 0.1 and 0.2 models are fairly simmilar with the 0.1 model slightly favoured, the total accretion and ratio are good (top and middle rows, green line) the instantaneous rate is slightly high but could be accounted for via a greater loss of mass to the intra cluster medium during a merger\footnote{The models presented here use $f_{loss}$ = 0.5 similar to \citet{Moster2018Emerge10}.}.
The models with $\gamma_z$ that give the closest fit to \citet{Mundy2017A3.5} are also closest to internal consistency. This has two implications: Firstly, an evolving high mass slope can create the central galaxy growth required for consistency with the satellite accretion predicted by \LCDM cosmology. Secondly, it suggests that the pair fraction can provide an additional constraint on the SMHM relation, in particular we see here the evolution of the pair fraction can constrain the evolution of the high mass slope.

\section{Discussion}
\label{sec:Discussion}

The primary goal of this paper is to show the propagation of systematics in galaxy modelling. 
Specifically, we use the SMHM relationship to connect assumptions used when estimating stellar masses from observations to systematics in galaxy pair fractions in the context of a \LCDM Universe. 
In this work we have used two observed stellar mass functions from SDSS-DR7 observations that use a de Vaucoulers and a S\`ersic + Exponential fit to determine stellar masses, which generate stellar mass functions with notably different number densities at the high mass end. Each stellar mass function generates, through abundance matching, a different SMHM relationship. The S\`ersic + Exponential mass function generates a steeper high-mass slope in the SMHM relationship at any epoch.
In addition to the SMHM relationships from the observed data we use a relationship fitted to match the outputs of the Illistris simulation. Furthermore, we also consider a toy model SMHM relation individually perturbing each input parameter to transparently probe the impact of the input SMHM relation on the pair fractions.
In each case we find that small changes introduced into the SMHM relationship can have significant effects on the expected pair fractions, as shown in Figures \ref{fig:MassRatioCartoon},\ref{fig:SMHM_PF_Cartoon}, \& \ref{fig:PairFracSystematic}.
This suggests that in the contxet of a \LCDM Universe tensions in previous observational studies could, in large part, be traced back to systematics in stellar mass estimates.

In \citet{Mundy2017A3.5} the $M_{*,cen}$ > $10^{10}$ pair fraction is given, however, this is not significantly different from the $M_{*,cen}$ > $10^{11}$ pair fraction shown in Figure \ref{fig:PairFractionData}. In Figure \ref{fig:PairFracSystematic} we find the pair fraction drops significantly when mass a selection is taken below the SMHM relation knee. As this drop is not found by \citet{Mundy2017A3.5} we interpret that their pair fraction measurement is not consistent with a break in the SMHM relation between $10^{10} M_{*,cen}$ and $10^{11} M_{*,cen}$.

\citet{Man2016RESOLVING03} noticed that the choice between luminosity-selected and stellar-mass selected pairs affected the pair fraction evolution. In this work we have provided a clear framework to properly interpret how input choices create systematic effects in the observed pair fraction and its evolution. Furthermore, it is a common approach to infer the assembly history of galaxies by converting the pair fractions into merger rates by assigning timescales to galaxy pairs \citep{Conselice2003A3,Conselice2008TheField,Mundy2017A3.5}. 
In \Paper{2} we developed a model that calculates the stellar mass growth rates of central galaxies and the stellar mass accretion rate from satellite galaxies.
It is found that, for some SMHM relationships, the accretion rate can be greater than the total growth rate implying the model is internally inconsistent and the SMHM relationship is not compatible with this \LCDM cosmology.
The stellar mass accretion rate is connected to the merger frequency and therefore the galaxy pair fraction. 
In this work we connect the shape and evolution of the SMHM relationship to the evolution of the pair fractions.
We propose it is therefore possible to use the pair fraction as an additional constraint to the SMHM relationship, this is a natural extension of conditional abundance matching or extended SHAM (subhalo abundance matching) models \citep{Hearin2013SHAMGroups}.
Using \steel one can test simultaneously the accretion ratio and the pair fraction generated from a given stellar mass function and cosmology.

Any changes to the stellar mass estimates such as photometry, background subtraction, IMF, e.t.c. that affect the stellar mass function will in a given cosmology create a change in the SMHM relationship. Therefore by the systematic propagation demonstrated in this work any stellar mass estimation will create systematic differences in the pair fractions. 
With the techniques presented in this work and in \Paper{1} \& \Paper{2}, one could retrieve the systematic differences created in pair fraction under multiple \LCDM cosmologies and for any given set of stellar mass functions. 
As a further test of our results we show the pair fractions found in Illustris, together with the pair fractions predicted by $\steel$, using a SMHM relationship designed to match
that of Illustris are fully consistent.
Using \steel the pair fractions produced from different \LCDM cosmologies can be tested to determine which best fits the observed pair fraction when ensuring self consistency in stellar mass estimates is maintained throughout.

In the era of wide and deep surveys, such as EUCLID, constraining a model using a single multi-epoch data set with consistent photometry will become a reality. The advantages of this are twofold: By tuning the SMHM relation to a given survey over a large range of redshifts the growth of the stellar mass function over time can be tested against the implied satellite accretion and star formation rate as in \Paper{2} this can be seen as a test of the consistency of the cosmological model or of the consistency of the stellar mass and/or starformation rate estimation. Secondly, as in \Paper{1} one can test if the high redshift SMHM relation produces the low redshift satellite distributions. The constraints on a given photometry, cosmological model, satellite evolution, starformation rate, e.t.c... are still not complete however it will allow nonphysical results to be identified. Furthermore, by making the model accessible it can then be used in the manner described above to make systematic adjustments to compare between current and future data sets that may use different stellar mass estimations.

\section{Conclusions}
\label{sec:Conclusions}

In this paper, we show that the input SMHM relations, based on different stellar mass estimations, have a significant impact on the predicted galaxy pair fraction. In short, the steeper the relation, the lower the pair fraction. Specifically we compare stellar mass functions created with a de Vaucoulers-based photometry (cmodel) to a S\'ersic-Exponential photometry (PyMorph), the latter leading to an enhancement in the number density of high mass galaxies. The resulting effect of these stellar mass functions is a different input SMHM relation to $\steel$, the primary difference consisting of a steeper high mass slope when adopting the S\'ersic-Exponential profile. As expected, the S\'ersic-Exponential results in a lower pair-fraction. To attempt to explain the difference in pair-fraction evolution with redshift we create a suite of toy models testing different alterations to the SMHM relation. We find that this evolution is linked to the evolution of the high mass slope.

The purpose of this paper is to show how subtle changes in derivation of stellar mass could lead to large differences in the pair fraction observed. This work is particularly important in the context of understanding galaxy mergers, the rate of which is commonly inferred from pair fractions, and the effects thereof. Merger rates are used to: predict rates of mass accretion, invoke variations in morphological type in many galaxy models, and are thought to relate to the triggering of AGN. It is therefore crucial when comparing two different samples to account for the systematic biases introduced by the assumptions implicit in stellar mass estimation as shown in this work. Future surveys should look to use fast and flexible modelling alongside data products to be able to properly understand the systematic effects of assumptions made on derived data products. For example the SMHM relation must simultaneously fit: traditional abundance matching, self-consistency between satellite accretion and central galaxy growth, and as shown in this paper the normalisation and evolution of the galaxy pair fraction. This multi-product fitting will ensure relations such as the SMHM relation are better constrained. The methods described in this work and \Paper{2} will also provide more stringent theoretical limits to the stellar mass estimations and photometerys used for future surveys.

\section*{Acknowledgements}
We acknowledge extensive use of the Python libraries astropy, matplotlib, numpy, pandas, and scipy. PJG acknowledges support from the STFC for funding this PhD. FS acknowledges partial support from a Leverhume Trust Research Fellowship.




\bibliographystyle{mnras}
\bibliography{STEEL3} 



\appendix


\bsp	
\label{lastpage}
\end{document}